\documentclass{moriond}

\usepackage{amsfonts}

\def\be{\begin{equation}}
\def\ee{\end{equation}}
\def\bea{\begin{eqnarray}}
\def\eea{\end{eqnarray}}

\newcommand{\Photo}{}

\begin{document}
\vspace*{4cm}
\title{Gravitational waves from compact binaries to the fourth post-Newtonian order}

\author{Luc Blanchet$^{a}$, Guillaume Faye$^{a,b}$, Quentin Henry$^{c}$, Fran\c{c}ois Larrouturou$^{d,\ast}$, David Trestini$^{a,e}$.}

\address{
$^{a)}$ 
${\mathcal{G}}{\mathbb{R}}\varepsilon{\mathbb{C}}{\mathcal{O}}$, Institut d'Astrophysique de Paris,\\
UMR 7095, CNRS, Sorbonne Universit{\'e}, 98\textsuperscript{bis} boulevard Arago, 75014 Paris, France.\\
$^{b)}$ 
Centre for Strings, Gravitation and Cosmology, Department of Physics,\\
Indian Institute of Technology Madras, Chennai 600036, India.\\
$^{c)}$
Max Planck Institute for Gravitational Physics\\
(Albert Einstein Institute), D-14476 Potsdam, Germany.\\
$^{d)}$
Deutsches Elektronen-Synchrotron DESY,\\
Notkestr. 85, 22607 Hamburg, Germany.\\
$^{e)}$
Laboratoire Univers et Th\'eories, Observatoire de Paris,\\
Universit\'e PSL, Universit\'e Paris Cit\'e, CNRS, F-92190 Meudon, France.}

\maketitle\abstracts{The precise knowledge of the gravitational phase evolution of compact binaries is crucial to the data analysis for gravitational waves. Until recently, it was known analytically (for non-spinning systems) up to the 3.5 post-Newtonian (PN) order, \emph{i.e.} up to the $(v/c)^7$ correction beyond the leading order quadrupole formula. Using a PN-multipolar-post-Minkowskian algorithm, we have pushed the accuracy to the next 4PN level. This derivation involved challenging technical issues, due to the appearance of non-physical divergences, which have to be properly regularized, as well as effects of non-linear multipole interactions.
}

\section{Motivations}

The signal analysis of gravitational-wave (GW) detectors strongly relies on analytic predictions, notably when it comes to observing the long inspiral phase of coalescing compact objects. In the current LIGO-Virgo-KAGRA network, those are typically binaries of neutron stars and, as such, a minority of the observed events. But despite their scarcity, such events encrypt important scientific information, and accurate waveform modeling is crucial to extract it. Moreover, the future generations of detectors (notably LISA or ET) are expected to observe a large amount of long inspiral signals. Establishing high-accuracy analytic waveform templates is thus a key goal for GW astronomy.

As of today, there exist two main approximation methods providing analytic waveforms for binaries of compact objects: the gravitational self-force framework, based on a small mass ratio expansion, and weak-field techniques. The post-Newtonian (PN) approach, on which this work relies, belongs to the second one. Indeed, it considers two objects in a gravitationally bound orbit, with $(v/c)^2 \sim Gm/(rc^2) \ll 1$, where $v$ is the typical velocity, $m = m_1 +m_2$ is the total mass and $r$ is the typical separation. As such, it is well tailored to explore the inspiral phase, when the two objects are significantly separated.

Our aim is to provide analytic predictions for the gravitational waveform, written in its WKB form as $h \sim A \,\mathrm{e}^{\mathrm{i}\psi}$. Here, we present the computation of the phase $\psi$ up to the 4.5PN accuracy, \emph{i.e.} the $(v/c)^9$ correction to the leading order. The results exposed hereafter are detailed in two recent publications~\cite{phase4PN,flux4PN}. As the detectors are more sensitive to the phase than to the amplitude $A$, we do not discuss the latter, but let the interested reader refer to~\cite{phase4PN,flux4PN} and references therein.

Note that we use a point-particle approximation, suitable for the inspiral phase of compact binaries, neglecting spins, finite-size effects or any ``physical'' properties of the compact objects other than mass. We let the reader refer to the proceedings of Q. Henry~\cite{QH} for the inclusion of spins, and of C. Aykroyd~\cite{CA} for the effect of the magnetic fields of the stars.

\section{Computing the gravitational flux and phase at 4PN}

\subsection{The balance equation}

The computation of the gravitational phase for binaries of compact objects, evolving in quasi-circular orbits, relies on the energy \emph{flux-balance} equation
\begin{equation}\label{eq:fluxbalance}
\frac{\mathrm{d} E}{\mathrm{d} t} = - \mathcal{F}\,,
\end{equation}
where $E$ is the conservative energy and $\mathcal{F}$ is the gravitational flux. On quasi-circular orbits, both quantities only depend on the frequency $\omega$, and Eq.~(\ref{eq:fluxbalance}) reduces to an easily solvable ordinary differential equation for the time evolution $\omega(t)$. Then, a further integration yields the gravitational phase. In order to compute it at a given PN order, one thus needs to know both $E$ and $\mathcal{F}$ at this same PN order.

The left-hand side of Eq.~(\ref{eq:fluxbalance}) is known at 4PN, and partial results exist for higher accuracies, see the proceedings of S. Foffa~\cite{SF} and references therein. We thus need to compute the flux at 4PN order. In the PN framework, the famous Einstein's quadrupole formula is extended as
\begin{equation}\label{eq:fluxThorne}
\mathcal{F} = \sum_{\ell \geq 2} \frac{G}{c^{2\ell+1}}\bigg[\frac{(\ell+1)(\ell+2)}{(\ell-1)\ell\,\ell!\,(2\ell+1)!!}\,\mathrm{U}_L^{(1)}\mathrm{U}_L^{(1)} + \frac{4\ell(\ell+2)}{(\ell-1)\,(\ell+1)!\,(2\ell+1)!!\,c^2}\,\mathrm{V}_L^{(1)}\mathrm{V}_L^{(1)}\bigg]\,,
\end{equation}
where parenthesis denote time derivatives, and two sets of \emph{radiative} moments $\{\mathrm{U}_L,\mathrm{V}_L\}$ are introduced. Those moments, of respectively mass- and current-type, are by definition those parametrising the metric at spatial infinity, in an appropriate transverse and traceless Bondi-like coordinate system~\cite{Thorne}. Thus, the computation of the flux at 4PN beyond the leading order boils down to the derivation of the mass quadrupole $\mathrm{U}_{ij}$ at 4PN, the mass octupole $\mathrm{U}_{ijk}$ and current quadrupole $\mathrm{V}_{ij}$ at 3PN, and so on.

\subsection{The mass quadrupole at 4PN}

Let us briefly overview the computation of the mass quadrupole at 4PN. More details on this derivation, as well as on the computation of the other needed moments, are to be found in~\cite{flux4PN} and references therein. Due to the non-linear nature of general relativity, the \emph{radiative} quadrupole $\mathrm{U}_{ij}$ can be split in two parts: (i) a \emph{source} quadrupole $\mathrm{I}_{ij}$, which is the moment defined in the near zone of the binary system; (ii) non-linear corrections, 
taking into account the various non-linear interactions experienced by the wave during its propagation:
\begin{equation}\label{eq:nonlinear}
\mathrm{U}_{ij} = 
\mathrm{I}_{ij}^{(2)} + \frac{1}{c^3}\,\mathrm{(tails)} + \frac{1}{c^5}\,\mathrm{(memory)} + \ldots  +\frac{1}{c^8}\,\mathrm{(tails-of-memory)} + \ldots \,.
\end{equation} 
Those can be classified according to their nature: \emph{tail} corresponds to simple scattering onto the static curvature of the space-time, \emph{memory} effects are associated with wave reemission by linear waves, \emph{etc}. At 4PN, a new interaction appears, which is cubic and dubbed \emph{tail-of-memory}, namely the scattering of the memory waves onto the static space-time curvature. Its intricate derivation can be found in~\cite{ToM1,ToM2}.

As for the source quadrupole $\mathrm{I}_{ij}$, its definition relies on a subtle matching procedure~\cite{matching} allowing to obtain it in terms of the near-zone metric. 
With this expression at hand, and once a model for the source has been chosen, the procedure is in principle straightforward. First, the near zone metric is obtained and injected into the explicit expression of $\mathrm{I}_{ij}$, and integrations by part are applied to reduce the tougher sectors. Then, the integrand is Taylor-expanded up to the desired (4PN) order, and integrated, using various techniques exposed in~\cite{MQ4PN_Had}. 

\subsection{Dealing with divergences}

The computation of the source quadrupole $\mathrm{I}_{ij}$ is plagued with two types of divergences. Naturally, those are unphysical and should not affect observable quantities. The first type is of ``UV'' nature, and appears due to our point-particle approximation. Indeed, we model the compact objects by Dirac distributions, and the infinite self-field of point particles has to be regularized. We have shown that the divergences, once properly treated by a dimensional regularization scheme, disappear from the mass quadrupole~\cite{MQ4PN_Had}.

On the other hand, divergences of the ``IR'' type are inherent to the PN framework. When defining the moments, we insert a formal PN expansion defined in the near zone and typically divergent at infinity, and integrate it up to infinity. A Hadamard regularization scheme was used in~\cite{MQ4PN_Had}, however, at 4PN, consistently with the 4PN equations of motion, we have to abandon it for a more powerful dimensional regularization.
This new IR scheme induces poles $\propto (d-3)^{-1}$, where $d$ is the number of spatial dimensions, which remain in the expression of the source quadrupole $\mathrm{I}_{ij}$. However, $\mathrm{I}_{ij}$ is not an observable quantity \emph{per se}, and those remaining divergences are exactly cancelled by other divergences coming from the non-linearities~\cite{MQ4PN_regIR,MQ4PN_renorm}, essentially \emph{tail-of-tail} and \emph{tail-of-memory}, in Eq.~(\ref{eq:nonlinear}). 

\section{Results and tests}

Collecting all the moments required by Eq.~(\ref{eq:fluxThorne}), we computed the gravitational energy flux at 4PN. To this result, one can add the 4.5PN coefficient, already known for quasi-circular orbits~\cite{45PN}. In terms of the PN frequency parameter $x = (Gm\omega/c^3)^{2/3}$, the 4.5PN flux reads
\bea\label{flux4.5PN}
	&& \mathcal{F} = \frac{32c^5}{5G}\nu^2 x^5 \Biggl\{
	1 
	+ \biggl(-\frac{1247}{336} - \frac{35}{12}\nu \biggr) x 
	+ 4\pi x^{3/2}
	\nonumber\\
	&& \quad
	+ \biggl(-\frac{44711}{9072} +\frac{9271}{504}\nu + \frac{65}{18} \nu^2\biggr) x^2 
	+ \biggl(-\frac{8191}{672}-\frac{583}{24}\nu\biggr)\pi x^{5/2}
	\nonumber\\
	&&  \quad
	+ \Biggl[\frac{6643739519}{69854400}+ \frac{16}{3}\pi^2-\frac{1712}{105}\gamma_\mathrm{E} - \frac{856}{105} \ln (16\,x) 
	+ \biggl(-\frac{134543}{7776} + \frac{41}{48}\pi^2 \biggr)\nu 
	- \frac{94403}{3024}\nu^2 
	- \frac{775}{324}\nu^3 \Biggr] x^3 
	\nonumber\\
	&& \quad
	+ \biggl(-\frac{16285}{504} + \frac{214745}{1728}\nu +\frac{193385}{3024}\nu^2\biggr)\pi x^{7/2} 
	\nonumber\\
	&& \quad
	+ \Biggl[ -\frac{323105549467}{3178375200} + \frac{232597}{4410}\gamma_\mathrm{E} - \frac{1369}{126} \pi^2 + \frac{39931}{294}\ln 2 - \frac{47385}{1568}\ln 3 + \frac{232597}{8820}\ln x   
	\nonumber\\
	&&  \quad\qquad
	+ \biggl( -\frac{1452202403629}{1466942400} + \frac{41478}{245}\gamma_\mathrm{E} - \frac{267127}{4608}\pi^2 + \frac{479062}{2205}\ln 2 + \frac{47385}{392}\ln 3  + \frac{20739}{245}\ln x \biggr)\nu
	\nonumber\\
	&&  \quad\qquad
	+ \biggl( \frac{1607125}{6804} - \frac{3157}{384}\pi^2 \biggr)\nu^2 + \frac{6875}{504}\nu^3 + \frac{5}{6}\nu^4 \Biggr] x^4
	\nonumber\\ 
	&&  \quad
	+ \Biggl[ \frac{265978667519}{745113600} - \frac{6848}{105}\gamma_\mathrm{E} - \frac{3424}{105} \ln (16 \,x)
	+ \biggl( \frac{2062241}{22176} + \frac{41}{12}\pi^2 \biggr)\nu
	\nonumber\\ 
	&&  \quad\qquad
	- \frac{133112905}{290304}\nu^2 - \frac{3719141}{38016}\nu^3 \Biggr] \pi x^{9/2} 
	+ \mathcal{O}\bigl(x^5\bigr) \Biggr\}\,,
\eea
where $\nu = m_1m_2/m^2$ is the symmetric mass ratio, $\gamma_\mathrm{E}$ is the Euler constant. The phase to be used in the data analysis of GW detectors is readily deduced from the balance equation~(\ref{eq:fluxbalance}), and numerical estimates suggest that the PN series converges well, at least up to 4.5PN order~\cite{phase4PN}.

\begin{figure}[h]
\begin{center}
\includegraphics[width=0.8\linewidth]{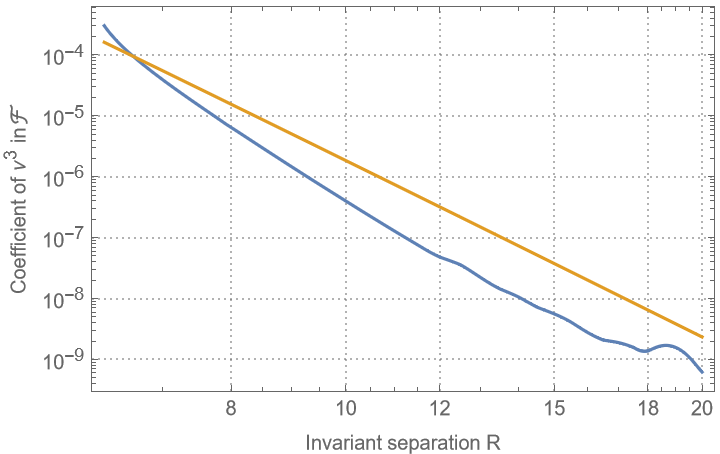}
\end{center}
\caption[]{Comparison of the analytic 4.5PN flux~(\ref{flux4.5PN}) and numerical second-order self-force results. The blue line is the difference between the two results, and the orange line is a typical 5PN behavior. It appears that the difference is compatible with a 5PN behavior, expressing a nice agreement up to 4.5PN. Courtesy to A. Pound.}
\label{fig:2GSF}
\end{figure}

Our result~(\ref{flux4.5PN}) was naturally subjected to various tests. In order to check the internal consistency of our derivation, we verified that all non-physical regularization scales disappear from observable quantities. On the other hand, we have also checked our computation against external results. First, when removing one of the two particles, our mass quadrupole $\mathrm{I}_{ij}$ reduces to the one of a boosted Schwarzschild black hole~\cite{BSS}. Then, when performing the small mass ratio limit $\nu \to 0$, all our results agree with first-order self-force analytic computations. Finally, our flux was also confronted against second-order self-force numerical results~\cite{2SF}, and passed it with flying colors~\cite{comp2SF}, as displayed in Fig.~\ref{fig:2GSF}.

\section*{Acknowledgments}

We thank Adam Pound for enlightening discussions during the 57$^\mathrm{th}$ \emph{Rencontres de Moriond}, and for providing us with the figure~\ref{fig:2GSF}. 
F.L. received funding from the European Research Council (ERC) under the European Union’s Horizon 2020 research and innovation program (grant agreement No 817791).


\section*{References}

\begin{thebibliography}{99}

\bibitem{phase4PN}L. Blanchet, G. Faye, Q. Henry, F. Larrouturou and D. Trestini, arXiv:2304.11185 (2023).

\bibitem{flux4PN}L. Blanchet, G. Faye, Q. Henry, F. Larrouturou and D. Trestini, arXiv:2304.11186 (2023).

\bibitem{QH}Q. Henry, {\em these Proceedings}, (2023).

\bibitem{CA}C. Aykroyd, {\em these Proceedings}, (2023).

\bibitem{SF}S. Foffa, {\em these Proceedings}, (2023).

\bibitem{Thorne}K. Thorne, {\em Rev. Mod. Phys.} \textbf{52}, 299 (1980).


\bibitem{ToM1}D. Trestini, F. Larrouturou, L. Blanchet, {\em Class. Quant. Grav.}, \textbf{40}, 055006 (2023).

\bibitem{ToM2}D. Trestini, L. Blanchet, {\em Phys. Rev. D}, to appear (2023).

\bibitem{matching}L. Blanchet, {\em Class. Quant. Grav.} \textbf{15}, 1971 (1998).

\bibitem{MQ4PN_Had}T. Marchand, Q. Henry, F. Larrouturou, S. Marsat, G. Faye and L. Blanchet, {\em Class. Quant. Grav.} \textbf{37}, 215006 (2020).

\bibitem{MQ4PN_regIR}F. Larrouturou, Q. Henry, L. Blanchet and G. Faye, {\em Class. Quant. Grav.} \textbf{39}, 115007 (2022).

\bibitem{MQ4PN_renorm}F. Larrouturou, L. Blanchet, Q. Henry and G. Faye, {\em Class. Quant. Grav.} \textbf{39}, 115008 (2022).

\bibitem{45PN}T. Marchand, L. Blanchet and G. Faye, {\em Class. Quant. Grav.} \textbf{33}, 244003 (2016).

\bibitem{BSS}L. Blanchet, T. Damour and B. Iyer, {\em Class. Quant. Grav.} \textbf{22}, 155 (2005).

\bibitem{2SF}N. Warburton, A. Pound, B. Wardell, J. Miller and L. Durkan, {\em Phys. Rev. Lett.} \textbf{127}, 15, 151102 (2021).

\bibitem{comp2SF}L. Blanchet, L. Durkan, G. Faye, Q. Henry, F. Larrouturou, J. Miller, A. Pound, D. Trestini, N. Warburton and B. Wardell, {\em in preparation}.
\end{thebibliography}
\end{document}